\documentclass{PoS}

\usepackage{amsmath,amssymb}
\usepackage{dsfont}
\usepackage{slashed}
\usepackage{verbatim}    
\usepackage{graphicx}

\newcommand{\be}{\begin{equation}}
\newcommand{\ee}{\end{equation}}
\newcommand{\bea}{\begin{eqnarray}} 
\newcommand{\eea}{\end{eqnarray}}

\newcommand{\qqq}{\displaystyle{\not}q_d}
\newcommand{\qq}{\displaystyle{\not}q_s}
\newcommand{\bpsi}{\overline{\psi}}
\newcommand{\Dr}{\displaystyle{\not}{\overrightarrow{D}}}
\newcommand{\Dl}{\displaystyle{\not}{\overleftarrow{D}}}
\newcommand{\partl}{\displaystyle{\not}{\overleftarrow{\partial}}}
\newcommand{\partr}{\displaystyle{\not}{\overrightarrow{\partial}}}
\newcommand{\DDr}{\overrightarrow{D}}
\newcommand{\DDl}{\overleftarrow{D}}
\newcommand{\MSbar}{{\overline{\rm MS}}}

\title{Perturbative and non-perturbative renormalization results of the Chromomagnetic Operator on the Lattice}

\ShortTitle{Renormalization results of the Chromomagnetic Operator on the Lattice}

\bigskip
\author{M. Constantinou$^a$, \speaker{M. Costa}\,$^a$, R. Frezzotti$^b$,
  V. Lubicz$^c$,
 G. Martinelli$^d$, D. Meloni$^c$,
  H.~Panagopoulos$^a$, S. Simula$^c$ \\
\llap{$^a$} Department of Physics, University of Cyprus, CY-1678 Nicosia,
  Cyprus\\
\llap{$^b$} Dipartimento di Fisica, Universit\`a di Roma ``Tor Vergata''
and INFN Sezione ``Tor Vergata'', I-00133 Rome, Italy\\
\llap{$^c$} Dipartimento di Fisica, Universit\`a Roma Tre, and INFN,
Sezione di Roma Tre, I-00146 Rome, Italy\\
\llap{$^d$} SISSA, I-34136 Trieste, Italy

\bigskip
E-mail: \email{constantinou.martha@ucy.ac.cy},
        \email{kosta.marios@ucy.ac.cy}, \email{roberto.frezzotti@roma2.infn.it},
        \email{lubicz@fis.uniroma3.it}, \email{guido.martinelli@sissa.it},
        \email{meloni@fis.uniroma3.it}, \email{haris@ucy.ac.cy}, \email{simula@roma3.infn.it}}

\abstract{The  Chromomagnetic operator (CMO) mixes with a large number of operators under renormalization. 
We identify which operators can mix with the CMO, at the quantum level. 
Even in dimensional regularization (DR), which has the simplest mixing
pattern, the CMO mixes with a total of 9 other operators, forming a basis of
dimension-five, Lorentz scalar 
operators with the same flavor content as the CMO. Among them, there
are also gauge noninvariant operators;
these are BRST invariant and vanish by the equations of motion, as required by
renormalization theory. On the other hand using a lattice regularization
further operators with $d \leq 5$ will mix; choosing the lattice action in a manner
as to preserve certain discrete symmetries, a minimul set of 3 additional operators (all with $d<5$) will appear.
In order to compute all relevant mixing coefficients, we calculate the quark-antiquark (2-pt) 
and the quark-antiquark-gluon (3-pt) Green's functions of the CMO at nonzero quark masses.
These calculations were performed in the continuum (dimensional regularization) and on the lattice
using the maximally twisted mass fermion action and the Symanzik improved gluon action. In parallel, non-perturbative measurements of the $K-\pi$ matrix element are being performed in simulations with 4 dynamical ($N_f = 2+1+1$) twisted mass fermions and the Iwasaki improved gluon action.}

\FullConference{32nd International Symposium on Lattice Field Theory LATTICE 2014\\
		 June 23 -- June 28, 2014\\
		 New York, USA}

\begin{document}

\section{Definition of the Chromomagnetic operator}

We study the mixing of the chromomagnetic operator (CMO):
{\small
\begin{equation}
{\cal O}_{CM}=g_0\,\bpsi_s\, \sigma_{\mu \nu}\, G_{\mu \nu} \psi_d,
\end{equation}}
\hspace{-0.1cm}using both dimensional regularization (DR) and lattice regularization (L).
The CMO appears naturally, via the OPE, in an effective Hamiltonian  description of semi-leptonic processes. For $\Delta S=1$ transitions, the low-energy effective Hamiltonian $H_{\rm eff}$ contains four magnetic operators of dimension 5:
{\small
\be
H^{\Delta S = 1,\ d=5}_{\rm eff} = \sum_{i=\pm} (C^i_\gamma(\mu) Q^i_\gamma(\mu) + C^i_g(\mu) Q^i_g(\mu)) + {\rm h.c.}
\ee
}
{\small
\be
Q_\gamma^\pm = {Q_d\,e\over 16 \pi^2} \left(\bpsi_{sL} \,\sigma^{\mu\nu}\, F_{\mu\nu} \, \psi_{dR} 
\pm \bpsi_{sR} \,\sigma^{\mu\nu}\, F_{\mu\nu} \, \psi_{dL}  \right), \,\, Q_g^\pm = { g\over 16 \pi^2} \left(\bpsi_{sL} \,\sigma^{\mu\nu}\, G_{\mu\nu} \, \psi_{dR}   
\pm \bpsi_{sR} \,\sigma^{\mu\nu}\, G_{\mu\nu} \, \psi_{dL} \right)
\ee
}
Some of the matrix elements of the CMO are parameterized as \cite{DIM} 
:
{\small
\bea 
\displaystyle\langle\pi^0|Q_g^+|K^0\rangle = {-11\over
  32\sqrt{2}\pi^2}\, {M_K^2 (p_\pi\cdot p_K)\over m_s+m_d}\, B_{g1},
\label{piQK}&&
\displaystyle\langle\pi^+\pi^-|Q_g^-|K^0\rangle =
        {11\,{\rm i}\over
  32\pi^2}\, {M_K^2\,M_\pi^2\over f_\pi\,( m_s+m_d)}\,  B_{g2}
\label{pipiQK}\\
\displaystyle\langle\pi^+\pi^+\pi^-|Q_g^+|K^+\rangle &=& {-11\over
  16\pi^2}\, {M_K^2\,M_\pi^2\over f_\pi^2\,( m_s+m_d)}\, B_{g3}.
\label{pipipiQK}
\eea
}
These matrix elements are relevant for the study of $K^0-\bar K^0$
mixing, $\epsilon^\prime/\epsilon$, the $\Delta I = 1/2$ rule, and
$K\to 3\pi$ decays. We focus on the matrix elements of ${\cal O}_{CM}$ 
between a kaon and a pion state. The $K-\pi$ matrix element of ${\cal O}_{CM}$  has 
never been calculated before on the lattice. The main difficulty of lattice calculation of the matrix elements of the CMO is that the strong interactions induce a mixing of the CMO with operators of lower dimension, with coefficients which are power divergent with the cutoff. The leading divergence of the bare CMO, which is of order $1/a^2$, is determined by the mixing with the dimension 3 scalar operator $\bpsi_s \psi_d$. Being power divergent, this subtraction is numerically difficult to control. In addition, its coefficient must be eventually evaluated in a fully non-perturbative way, since non-perturbative effects, e.g., factors of the form $a \Lambda_{QCD}$, 
combined with factors which diverge as inverse powers of the lattice spacing can give finite contributions~\cite{Maiani:1992vl}.

\section{Symmetries of the Action and Transformation Properties of operators}

With our choice of lattice action, there exist certain symmetries of the action (valid both in the
continuum and lattice formulation of the theory) which reduce
considerably the number of operators that can possibly mix with ${\cal
O}_{CM}$ at the quantum level. These symmetries are defined by means of the 
discrete transformations ${\cal{P}}$ (continuum parity),
${\cal{D}}_d$,
${\cal{R}}_5$ \cite{Frezzotti:2004},
${\mathcal {C}}$ (charge conjugation),
and $\cal{S}$ (exchange between the $s$ and the $d$ quark). In terms of the above transformations, the symmetries  of the action are:
{\small
\bea
&\bullet& {\cal{P}}\times {\cal D}_d \times (m \to - m){\rm,\, where\,} m\, {\rm are \, all\, masses\, except}\, M_{\rm{cr}} \nonumber\\
&\bullet& {\cal{D}}_d\times {\cal R}_5\nonumber\\
\label{symmetries}
&\bullet& {\cal{C}}\times{\cal{S}}, {\rm if}\,\, r_s=r_d\\\nonumber
&\bullet& {\cal{C}}\times{\cal{P}}\times{\cal{S}}, {\rm if} \,\,r_s=-r_d\,.\nonumber
\eea}
In order to identify which operators can possibly mix with ${\cal O}_{CM}$, we examine the transformation properties of all candidate operators under the above symmetries; admissible operators must transform in the same way as ${\cal O}_{CM}$. Furthermore, by general renormalization theorems, these operators must be gauge invariant, or else they must vanish by the equations of motion (eom).
{
{\tiny
\begin{table}[h!]
\begin{center}
\begin{tabular}{| c|  p{6.5cm} | c| c| c | c|}
\hline
\multicolumn{2}{|l|}{Operators}  &
${\cal{P}}\times {\cal D}_d \times$ & ${\cal{D}}_d\times {\cal R}_5$ & ${\cal{C}}\times{\cal{S}}$ & 
${\cal{C}}\times{\cal{P}}\times{\cal{S}}$\\
\multicolumn{2}{|l|}{}  & $(m \to - m)$ & $\,$ & if\,\,$r_s=r_d$& if\,\,$r_s=-r_d$\\\hline
\multicolumn{3}{l}{Dimension 3 operators}\\\hline  

$\,\checkmark$& {\small $\,\bpsi_s\psi_d$}  & $-$ & $+$ & $+$ & $+$\\[0.01ex] 

$\,$& {\small $\,i\,\bpsi_s\gamma_5\psi_d$}  & $+$ & $+$ & $+$ & $-$\\\hline 

\multicolumn{3}{l}{Dimension 4 operators}\\\hline 

$\,$& {\small $\,(m_{d}+m_{s})\bpsi_s\psi_d$} & $+$ & $+$ & $+$ & $+$\\[0.01ex] 

$\,$& {\small $\,(m_{d}-m_{s})\bpsi_s\psi_d$} & $+$ & $+$ & $-$ & $-$\\[0.01ex] 

$\,(+)$& {\small $\,i\,(m_{d}+m_{s})\bpsi_s\gamma_5\psi_d$} & $-$ & $+$
& $+$ & $-$\\[0.01ex] 

$\,(-)$& {\small $\,i\,(m_{d}-m_{s})\bpsi_s\gamma_5\psi_d$} & $-$ &
$+$ & $-$ & $+$ \\[0.01ex] 

$\,$& {\small $\,\bpsi_s(\Dr+m_d)\psi_d
+\bpsi_s(-\Dl +m_s)\psi_d$}&
$+$ & $+$ & $+$ & $+$ \\[0.01ex] 

$\,$& {\small $\,\bpsi_s(\Dr+m_d)\psi_d
-\bpsi_s(-\Dl +m_s)\psi_d$}
& $+$ & $+$ & $-$ & $-$\\[0.01ex] 

$\,(+)$& {\small $\,i\,\bpsi_s\gamma_5(\Dr+m_d)\psi_d
  +i\,\bpsi_s(-\Dl +m_s)\gamma_5\psi_d$}&
$-$ & $+$ & $+$ & $-$ \\[0.01ex] 

$\,(-)$& {\small $\,i\,\bpsi_s\gamma_5(\Dr+m_d)\psi_d
  -i\,\bpsi_s(-\Dl +m_s)\gamma_5\psi_d$} &
$-$ & $+$ & $-$ & $+$ \\\hline
\end{tabular}
\caption[Transformation properties of dimension 3 and 4 operators.]{Transformation properties of dimension 3 and 4 operators. Gauge non-invariant operators
do not appear, since they do not vanish by the eom.
}
\label{tb:GaugeVariantAnde.m.o1}
\end{center}
\end{table}}
}
In Tables~\ref{tb:GaugeVariantAnde.m.o1} and~\ref{tb:GaugeVariantAnde.m.o2} we present all candidate
operators along with their transformation properties. Operators marked
by $''\checkmark''$ have the same properties as ${\cal O}_{CM}$ and thus
may mix with it. Operators marked by $''(+)''$ or $''(-)''$ have the same
transformation properties as ${\cal O}_{CM}$ only if $r_s=r_d$ or
$r_s=-r_d$, respectively; for this reason the Wilson parameters
$r_s,\,r_d$ have been explicitly introduced in ${\cal O}_{11}$ and
${\cal O}_{12}$ below (see Eqs.~(\ref{O11}) - (\ref{O12})). There follows immediately that ${\cal
 O}_{CM}\equiv {\cal O}_1$ can only mix with the following operators:
{
\tiny{
\begin{table}
\begin{center}
\begin{tabular}{| c|  p{7.6cm} | c| c| c | c|}
\hline
\multicolumn{2}{|l|}{Dimension 5 Operators}  &
${\cal{P}}\times {\cal D}_d \times$ & ${\cal{D}}_d\times {\cal R}_5$ & ${\cal{C}}\times{\cal{S}}$ & 
${\cal{C}}\times{\cal{P}}\times{\cal{S}}$\\
\multicolumn{2}{|l|}{}  & $(m \to - m)$ & $\,$ & if\,\,$r_s=r_d$& if\,\,$r_s=-r_d$\\\hline
$\,\checkmark$&{\small$\,g_0\,\bpsi_s \sigma_{\mu \nu} G_{\mu \nu} \psi_d$}
& $-$ & $+$ & $+$ & $+$ \\[0.01ex] 

$\,$& {\small$\,i\,g_0\,\bpsi_s \gamma_5\sigma_{\mu \nu} G_{\mu \nu} \psi_d$}  & $+$
& $+$ & $+$ & $-$\\[0.01ex] 

$\,\checkmark$&{\small$\,(m_{d}^2+m_{s}^2)\bpsi_s\psi_d$} & $-$ & $+$ & $+$
& $+$ \\[0.01ex] 

$\,$& {\small$\,i\,(m_{d}^2+m_{s}^2)\bpsi_s\gamma_5\psi_d$} & $+$ & $+$ & $+$ & $-$
\\[0.01ex] 

$\,$& {\small$\,(m_{d}^2-m_{s}^2)\bpsi_s\psi_d$} & $-$ & $+$ & $-$ & $-$
\\[0.01ex] 

$\,$& {\small$\,i\,(m_{d}^2-m_{s}^2)\bpsi_s\gamma_5\psi_d$} & $+$ & $+$ & $-$ &
$+$\\[0.01ex] 

$\,\checkmark$&{\small$\,m_{d}\,m_{s}\bpsi_s\psi_d$} & $-$ & $+$ & $+$ & $+$
\\[0.01ex] 

$\,$& {\small$\,i\,m_{d}\,m_{s}\bpsi_s\gamma_5\psi_d$} & $+$ & $+$ & $+$ & $-$ \\[0.01ex] 

$\,\checkmark$&{\small$\,m_{s}\bpsi_s(\Dr+m_d)\psi_d
  +
  m_{d}\bpsi_s(-\Dl +m_s)\psi_d$}&
$-$ & $+$ & $+$ & $+$ \\[0.01ex] 

$\,\checkmark$&{\small$\,m_{d}\bpsi_s(\Dr+m_d)\psi_d
  +
  m_{s}\bpsi_s(-\Dl +m_s)\psi_d$}
& $-$ & $+$ & $+$ & $+$\\[0.01ex] 

$\,$& {\small$\,m_{s}\bpsi_s(\Dr+m_d)\psi_d
-
m_{d}\bpsi_s(-\Dl +m_s)\psi_d$}
& $-$ & $+$ & $-$ & $-$\\[0.01ex] 

$\,$& $\,m_{d}\bpsi_s(\Dr+m_d)\psi_d - 
m_{s}\bpsi_s(-\Dl +m_s)\psi_d$ & $-$ & $+$ & $-$ & $-$\\[0.01ex] 

$\,$& {\small$\,i\,m_{s}\bpsi_s\gamma_5(\Dr+m_d)\psi_d + 
i\,m_{d}\bpsi_s(-\Dl +m_s)\gamma_5\psi_d$} & $+$  & $+$ & $+$ & $-$\\[0.01ex] 

$\,$& {\small$\,i\,m_{d}\bpsi_s\gamma_5(\Dr+m_d)\psi_d + 
i\,m_{s}\bpsi_s(-\Dl +m_s)\gamma_5\psi_d$} & $+$  & $+$ & $+$ & $-$\\[0.01ex] 

$\,$& {\small$\,i\,m_{s}\bpsi_s\gamma_5(\Dr+m_d)\psi_d - 
i\,m_{d}\bpsi_s(-\Dl +m_s)\gamma_5\psi_d$}& $+$  & $+$ & $-$ & $+$ \\[0.01ex] 

$\,$& {\small$\,i\,m_{d}\bpsi_s\gamma_5(\Dr+m_d)\psi_d - 
i\,m_{s}\bpsi_s(-\Dl +m_s)\gamma_5\psi_d$} & $+$  & $+$ &  $-$ & $+$\\[0.01ex] 

$\,\checkmark$&{\small$\,\bpsi_s(\Dr+m_d)^2\psi_d + 
\bpsi_s(-\Dl +m_s)^2\psi_d$}& $-$ & $+$ & $+$ & $+$ \\[0.01ex] 

$\,$& {\small$\,\bpsi_s(\Dr+m_d)^2\psi_d - 
\bpsi_s(-\Dl +m_s)^2\psi_d$} & $-$ & $+$ & $-$ & $-$\\[0.01ex] 

$\,$& {\small$\,i\,\bpsi_s\gamma_5(\Dr+m_d)^2\psi_d + 
i\,\bpsi_s(-\Dl +m_s)^2\gamma_5\psi_d$} & $+$ & $+$ & $+$ & $-$\\[0.01ex] 

$\,$& {\small$\,i\,\bpsi_s\gamma_5(\Dr+m_d)^2\psi_d - 
i\,\bpsi_s(-\Dl +m_s)^2\gamma_5\psi_d$}& $+$ & $+$ & $-$ & $+$ \\[0.01ex] 

$\,\checkmark$&{\small$\,\bpsi_s \DDl_\mu \DDr _\mu \psi_d$}& $-$ & $+$ & $+$ & $+$ \\[0.01ex] 

$\,$& {\small$\,i\,\bpsi_s \gamma_5\DDl_\mu \DDr _\mu \psi_d$}& $+$ & $+$ & $+$ & $-$ \\[0.01ex] 

$\,\checkmark$&{\small$\,\bpsi_s(-\Dl +m_s)\,(\Dr+m_d)\psi_d$}& $-$ & $+$ & $+$ & $+$ \\[0.01ex] 

$\,$& {\small$\,i\,\bpsi_s(-\Dl +m_s)\,\gamma_5\,(\Dr+m_d)\psi_d$}& $+$ & $+$ & $+$ & $-$ \\[0.01ex] 

$\,\checkmark$&{\small$\,\bpsi_s  \partl(\Dr +m_d )\psi_d 
  - \bpsi_s (-\Dl +m_s)\partr \psi_d $}& $-$ & $+$ & $+$ & $+$ \\[0.01ex] 

$\,\checkmark$&{\small$\,\bpsi_s  \partr (\Dr +m_d )\psi_d 
  - \bpsi_s (-\Dl +m_s)\partl\psi_d $}& $-$ & $+$ & $+$ & $+$ \\[0.01ex] 

$\,$& {\small$\,\bpsi_s \partl(\Dr +m_d )\psi_d 
  + \bpsi_s (-\Dl +m_s)\partr \psi_d $}& $-$ & $+$ & $-$ & $-$ \\[0.01ex] 

$\,$& {\small$\,\bpsi_s  \partr (\Dr  +m_d )\psi_d 
  + \bpsi_s (-\Dl +m_s)\partl\psi_d $}& $-$ & $+$ & $-$ & $-$ \\[0.01ex] 

$\,$& {\small$\,i\,\bpsi_s  \partl\gamma_5(\Dr   +m_d )\psi_d 
  - i\,\bpsi_s (-\Dl +m_s)\gamma_5\partr \psi_d $}& $+$ & $+$ & $+$ & $-$ \\[0.01ex] 

$\,$& {\small$\,i\,\bpsi_s \partr \gamma_5(\Dr +m_d )\psi_d 
  - i\,\bpsi_s (-\Dl +m_s)\gamma_5\partl\psi_d $}& $+$ & $+$ & $+$ & $-$ \\[0.01ex] 

$\,$& {\small$\,i\,\bpsi_s \partl\gamma_5(\Dr +m_d )\psi_d 
  + i\,\bpsi_s (-\Dl +m_s)\gamma_5\partr \psi_d $} & $+$ & $+$ & $-$ & $+$\\[0.01ex] 

$\,$& {\small$\,i\,\bpsi_s \partr \gamma_5(\Dr +m_d )\psi_d
  + i\,\bpsi_s (-\Dl +m_s)\gamma_5\partl\psi_d $}& $+$ & $+$ & $-$ & $+$ \\\hline
\end{tabular}
\caption[Transformation properties of dimension 5 operators]{Transformation properties of gauge invariant operators and of
  operators which vanish by the equations of motion, in the physical
  basis (Operator dimension = 5).}
\label{tb:GaugeVariantAnde.m.o2}
\end{center}
\end{table}}
}
{\small
\bea
{\cal O}_1    &=& g_0\,\bpsi_s \sigma_{\mu \nu} G_{\mu \nu}
\psi_d,  \quad
{\cal O}_2    = (m_{d}^2+m_{s}^2)\bpsi_s\psi_d, \quad
{\cal O}_3    =  m_{d}\,m_{s}\bpsi_s\psi_d, \quad
{\cal O}_4    = \Box(\bpsi_s \psi_d) \, \\[0.01ex]
{\cal O}_5    &=&
\bpsi_s(-\Dl +m_s)(\Dr+m_d)\psi_d,
\quad {\cal O}_6    =
\bpsi_s(\Dr+m_d)^2\psi_d
+
\bpsi_s(-\Dl +m_s)^2\psi_d \,
\\[0.01ex]
{\cal O}_7    &=&
m_{s}\bpsi_s(\Dr+m_d)\psi_d
+
m_{d}\bpsi_s(-\Dl +m_s)\psi_d,
\,\,\,
{\cal O}_8    =
m_{d}\bpsi_s(\Dr+m_d)\psi_d
+
m_{s}\bpsi_s(-\Dl +m_s)\psi_d\,
\\[0.01ex]
{\cal O}_9    &=&  \bpsi_s
\partl(\Dr+m_d
)\psi_d
                   - \bpsi_s
                   (-\Dl +m_s)\partr \psi_d,
                   \,\,\,
{\cal O}_{10} = \bpsi_s
\partr (\Dr+m_d
)\psi_d
                   - \bpsi_s
                   (-\Dl +m_s)\partl\psi_d\,\\[0.01ex]
\label{O11}
{\cal O}_{11} &=&
i\,r_d\,\bpsi_s\gamma_5(\Dr+m_d)\psi_d
+i\,r_s\,\bpsi_s(-\Dl +m_s)\gamma_5\psi_d,
\quad \label{O12}
{\cal O}_{12} =
i\,(r_d\,m_{d}+r_s\,m_{s})\bpsi_s\gamma_5\psi_d  \\[0.01ex]
{\cal O}_{13} &=& \bpsi_s\,\psi_d\,.
\eea
}
For the parameters $r_s$\,, $r_d$\,, in our perturbative calculation we have made the 
(independent) choices of values $r_s=\pm 1$,\, $r_d=\pm 1$, consistently with
their values in simulations. Operators ${\cal O}_{9}$ and ${\cal O}_{10}$ are not gauge invariant, but they vanish by
the eom; indeed, they will mix with ${\cal O}_{CM}$ 
both in dimensional regularization and on the lattice.
The operators ${\cal O}_{11},\,{\cal O}_{12},\,{\cal O}_{13}$ are of
lower dimension and thus they do not mix with ${\cal O}_{1}$ in
dimensional regularization; they do however show up in the lattice
formulation. The renormalized operators ${\cal O}^R_i$ are related to the bare ones,
${\cal O}_i$ ($i=1,\ldots,13$), through:
{\small
\be
{\cal O}_i=  \sum_{j=1}^{13} Z_{ij} {\cal O}_j^R\,\,\quad ({\rm in\,\, matrix\,\, notation:}\,\, {\cal O}=Z {\cal O}^R )\,,
\ee
}
\vspace{-0.5cm}
where $Z_{ij}$  is a $13 \times 13$ mixing matrix. For ${\cal O}_1^R$,
we only need the first row of $Z^{-1}$.
The bare amputated Green's functions of
${\cal O}_1$\, are related to the corresponding renormalized ones through:
{\small
\be
\langle \psi^R\,{\cal O}_1^R\,\bpsi^R \rangle_{\rm amp} = 
Z_\psi\,\sum_{i=1}^{13}(Z^{-1})_{1i} \langle\psi\,{\cal O}_i\,\bpsi \rangle _{\rm amp}\,,\quad \psi=\sqrt{Z_\psi}\,\psi^R
\label{GG2}
\ee
\be
\langle \psi^R\,{\cal O}_1^R\,\bpsi^R A_\nu^R \rangle_{\rm amp} =
Z_\psi\,Z_A^{1/2}\sum_{i=1}^{13}(Z^{-1})_{1i} \langle\psi\,{\cal
  O}_i\,\bpsi\,A_\nu \rangle _{\rm amp}\,,\quad  A_{\nu}= \sqrt{Z_A}\,A^{R}_{\nu}\,
\label{GG3}
\ee
}
\section{Lattice regularization -- Renormalization functions in the $\MSbar$ scheme}
The general form of the divergent (logarithmic) part of the bare Green's functions is:
{\small
\bea
\langle \psi\,{\cal O}_1\,\bpsi \rangle^{L}_{\rm amp}\Big{|}_{\log(a^2)} &=&\rho_1\,(q^2_s {+} q^2_d)
+ \rho_2\,(m_s^2 {+} m^2_d) + \rho_3\,i\,(m_d\qqq {+} m_s\qq) + \rho_4\,i\,(m_s\qqq {+} m_d\qq)\nonumber \\  
&&+ \rho_5\,q_s{\cdot}q_d  + \rho_6\,\qq\,\qqq + \rho_7\,m_sm_d + \rho_8\,(r_d\,\gamma_5\,\qqq {+} r_s\,\qq\,\gamma_5) \nonumber \\ 
&&+ \rho_9\,i\,(r_dm_d {+} r_sm_s)\,\gamma_5 + \rho_{10}\cdot 1 
\label{G2}
\eea
}
\vspace{-0.5cm}
{\small
\bea
\langle \psi\,{\cal O}_1\,\bpsi A_\nu \rangle^{L}_{{\rm amp}, 1PI}\Big{|}_{\log(a^2)} &=& R_1\,g
\,(q_s{+}q_d)_\nu + R_2\,g\,(\gamma_\nu\qqq+ \qq\gamma_\nu)
+ R_3\,i\,g\,(m_s{+}m_d)\gamma_\nu \nonumber \\
&&+ R_4\,(-2i\,g\,\sigma_{\rho\nu}q_{A\rho}) 
+  R_5\,g\,(r_d - r_s)\,\gamma_5\,\gamma_\nu 
\label{G3}
\eea
}
where $\rho_i,\,R_i$ are numerical coefficients and $q_s/q_d/q_A$ is the momentum of the external antiquark/quark/gluon. 
The $\MSbar$-renormalized Green's functions are already known in dimensional regularization, see Ref.~\cite{HP2013}. Equations~(\ref{GG2}) and (\ref{GG3}) now take the form:
{\small
\be
\langle \psi\,{\cal O}_1\,\bpsi \rangle_{\rm amp}^{\MSbar} = 
Z_\psi^{L,\MSbar}\,\sum_{i=1}^{13}\big((Z^{L,\MSbar})^{-1}\big)_{1i} \langle\psi\,{\cal O}_i\,\bpsi \rangle^{L}_{\rm amp},
\label{GG2lattice}
\ee
\be
\langle \psi\,{\cal O}_1\,\bpsi A_\nu \rangle_{\rm amp}^{\MSbar} =
Z_\psi^{L,\MSbar}\,(Z^{L,\MSbar}_A)^{1/2}\sum_{i=1}^{13}\big((Z^{L,\MSbar})^{-1}\big)_{1i}  \langle\psi\,{\cal
  O}_i\,\bpsi\,A_\nu \rangle^{L}_{\rm amp}\,.
\label{GG3lattice}
\ee}
Renormalizability of the theory implies that the difference between the one-loop renormalized 
and bare Green's functions must only consist of expressions which are polynomial in $q_i,\, m$ having the form shown in Eqs.~(\ref{G2}) - (\ref{G3});
in this way, the right-hand sides of Eqs.~(\ref{GG2lattice}) - (\ref{GG3lattice}) can be rendered
equal to the corresponding left-hand sides, by an appropriate definition of the ($q_i$- and $m$-independent)
renormalization functions $Z_{1i}^{L,\MSbar} \equiv Z_i^{L,\MSbar}$.
These differences can be written as follows:
{\small
\be
\label{RminusLattice2pt}
\langle \psi\,{\cal O}_1\,\bpsi \rangle_{\rm amp}^{\MSbar}-\langle \psi\,{\cal O}_1\,\bpsi \rangle_{\rm amp}^{L}=
g^2\Big(z_\psi^{L,\MSbar}+ z_g^{L,\MSbar}-z_1^{L,\MSbar}\Big) \langle \psi \,{\cal O}_1\,\bpsi \rangle_{\rm tree} - \sum_{i=2}^{13}Z^{L,\MSbar}_i  \langle\psi\,{\cal O}_i\,\bpsi \rangle _{\rm tree}
\ee
}
{\small
\be
\label{RminusLattice3pt}
\langle \psi\,{\cal O}_1\,\bpsi A_\nu \rangle_{\rm amp}^{\MSbar} - \langle \psi\,{\cal O}_1\,\bpsi A_\nu \rangle_{\rm amp}^{L}=g^2
\Big(z_\psi^{L,\MSbar}+\frac{1}{2}z_A^{L,\MSbar}+ z_g^{L,\MSbar} - z_1^{L,\MSbar} \Big) \langle \psi \,{\cal O}_1\,\bpsi A_\nu \rangle _{\rm tree}- \sum_{i=2}^{13}Z^{L,\MSbar}_i \langle\psi\,{\cal O}_i\,\bpsi\,A_\nu \rangle _{\rm tree}\,,
\ee
}
where $Z^{L,\MSbar}_1= 1+ g^2\,z^{L,\MSbar}_1 + {\cal O}(g^4)$, $Z^{L,\MSbar}_{i>1} = {\cal O}(g^2)$ and $Z^{L,\MSbar}_\psi=1+g^2\,z^{L,\MSbar}_\psi+{\cal O}(g^4)$\,,
$Z^{L,\MSbar}_A=1+g^2\,z^{L,\MSbar}_A+{\cal O}(g^4)\,,\quad Z^{L,\MSbar}_g=1+g^2\,z^{L,\MSbar}_g+{\cal O}(g^4)$\,.

Equations~(\ref{RminusLattice2pt}) and~(\ref{RminusLattice3pt}) amount to $10+5$
conditions for the $13$ mixing coefficients; they are consistent and admit a unique solution. Upon solving them we obtain for the Iwasaki gluon action:
{\small
\be
\label{Z1LMS}
Z^{L,\MSbar}_1  = 1+ \frac{g^2}{16\,\pi^2}\,
                     \Bigg(N_c \left(-7.9438 +\frac{1}{2}\,\log\left(a^2\,\bar\mu^2 \right) \right) + \frac{1}{N_c}  \left(4.4851  -\frac{5}{2}\,\log\left(a^2\,\bar\mu^2 \right) \right) \Bigg) 
\ee
}             
{\small
\bea
Z^{L,\MSbar}_2  &=& \frac{g^2\,C_F}{16\,\pi^2}\,\left(4.5370 + 6\,\log\left(a^2\,\bar\mu^2 \right)\right),
\quad Z^{L,\MSbar}_3  = Z^{L,\MSbar}_4  = 0
\\[0.01ex]
Z^{L,\MSbar}_5  &=&  \frac{g^2}{16\,\pi^2}\,
                   \left(N_c \left(4.2758 -\frac{3}{2}\, \log\left(a^2\,\bar\mu^2 \right)\right) + 
                         \frac{1}{N_c}  \Bigl(-3.7777 + 3\,\log\left(a^2\,\bar\mu^2 \right)\Bigr) \right)
\\[0.01ex]
Z^{L,\MSbar}_6  &=& 0, \quad
Z^{L,\MSbar}_7  = -\frac{Z^{L,\MSbar}_5}{2}, \quad Z^{L,\MSbar}_8  = \frac{g^2\,C_F}{16\,\pi^2}\,\left(-3.7760\right) 
\\[0.01ex]
Z^{L,\MSbar}_9  &=& \frac{Z^{L,\MSbar}_5}{2},
\quad
Z^{L,\MSbar}_{10} = \frac{g^2\,C_F}{16\,\pi^2}\,\left(3.7777 - 3\,\log\left(a^2\,\bar\mu^2 \right)\right) 
\\[0.01ex]
Z^{L,\MSbar}_{11} &=& \frac{1}{a}\,\frac{g^2\,C_F}{16\,\pi^2}\,\left(-3.2020\right),
\quad
Z^{L,\MSbar}_{12} = -Z^{L,\MSbar}_{11},
\quad
\label{Z13LMS}
Z^{L,\MSbar}_{13} = \frac{1}{a^2}\frac{g^2\,C_F}{16\,\pi^2}\,\left(36.0613\right)\,.
\eea
}
For the operators of lower dimensionality (${\cal O}_{11}$ -  ${\cal O}_{13}$), given that their
coefficients are power divergent, perturbation theory is expected
to provide only a ballpark estimate at best. Fortunately, it is
precisely for the coefficients of these latter operators that we can have best access to
non-perturbative estimates. 
\section{Non-perturbative results}
\begin{figure}[h!]
\begin{center}
\centering{\includegraphics[height=6cm]{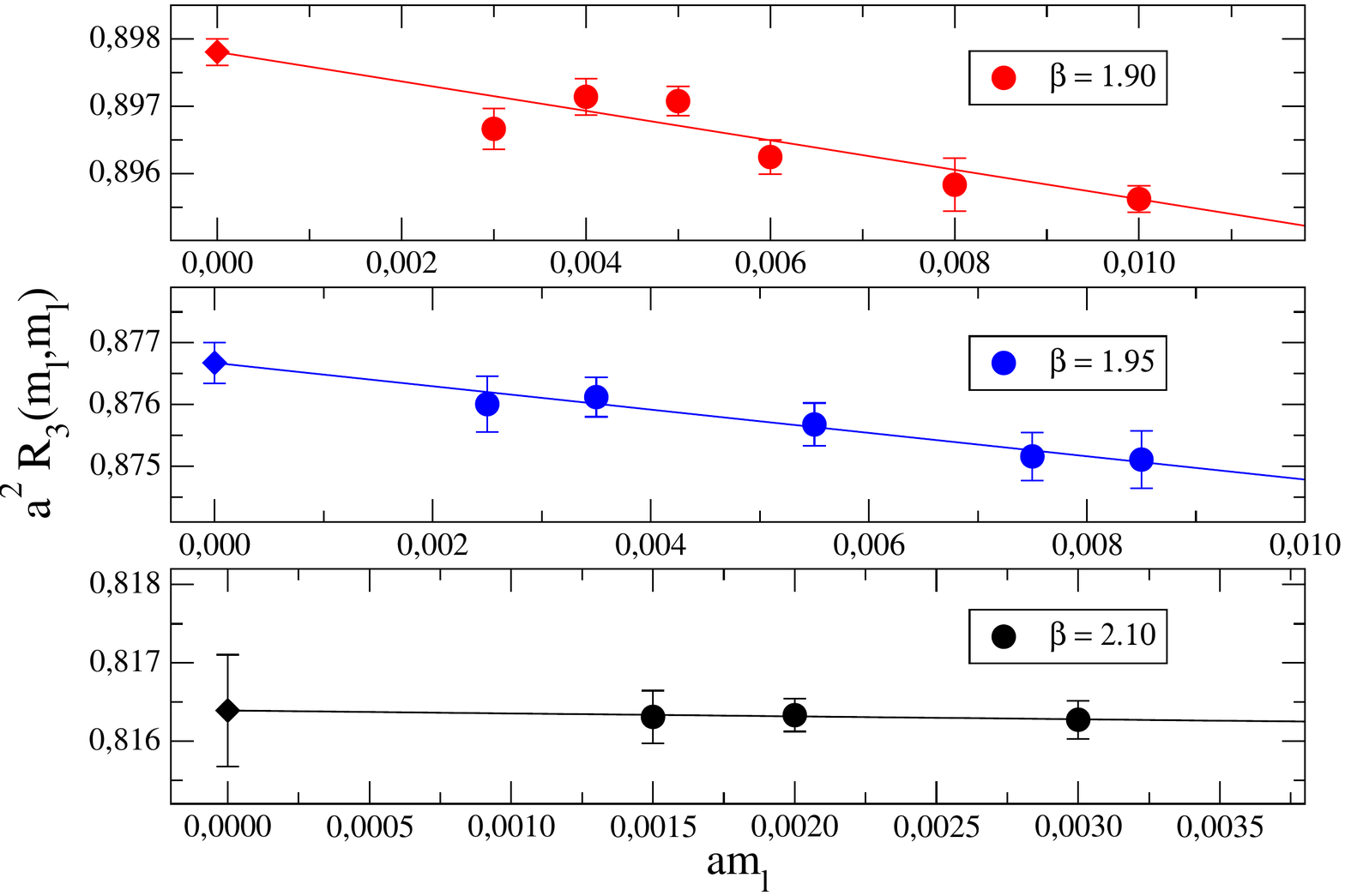}}
\caption{The ratio $a^2 R_3(m_l, m_l)$ versus the (twisted) quark bare mass $a m_l$ for each value of the lattice spacing. The solid lines are the results of linear fits in $a m_l$ applied to the data. The diamonds represent the values of the mixing coeffcient $c_{13}$ obtained in the chiral limit (after Ref.~\cite{VL2014}).
}
\label{fig3c13}
\end{center}
\end{figure}

In this Section we present our non-perturbative determination of the coefficient $Z_{13}$ describing the power-divergent
mixing of the chromomagnetic operator with the scalar density (${\cal O}_{13}$). We use lattice QCD simulations with the gauge configurations
produced by ETMC with four flavors of dynamical quarks ($N_f = 2 + 1 + 1$), which include in the sea,
besides two light mass degenerate quarks, also the strange and charm quarks with masses close to their physical values.
We impose non-perturbative conditions:
{\small
\be
\lim_{m_s\,,\ m_d \to 0} \langle\pi(0)|{\cal O}_1^{\rm
  R}|K(0)\rangle = \lim_{m_s\,,\ m_d \to 0} \langle\pi(0)|{\cal O}_1
  - \frac{c_{13}}{a^2}{\cal O}_{13}|K(0)\rangle = 0
\ee
\be
\langle 0|{\cal O}_1^{\rm R}|K(0)\rangle_{m_s\,,\ m_d} = \langle 0|{\cal O}_1
- \frac{c_{13}}{a^2}{\cal O}_{13} 
- \frac{c_{12}}{a}{\cal O}_{12}|K(0)\rangle_{m_s\,,\ m_d} = 0.
\label{c12}
\ee
}
In order to extract the coefficients $c_{13}$ and $c_{12}$ from these conditions we have computed for each ensemble the 2- and 3-pt meson correlators.
Note that the non-perturbative determination of $c_{12}$, based on Eq.~(\ref{c12}) is in progress. From the large time behavior of the 3-point correlators corresponding to the chromomagnetic and scalar density insertions one can compute $c_{13}$ in lattice units. In particular, from 
the following ratio:
\be
R_3(m_s,m_l) \equiv \frac{\langle \pi |{\cal O}_1|K\rangle}{\langle \pi |{\cal O}_{13}|K\rangle} \quad {\rm and} \quad c_{13} = \lim_{m_s,m_l \to 0} a^2R_3(m_s,m_l)
\ee
we can obtain an estimate of the mixing coefficient $c_{13}$ at each value of the lattice spacing. The results obtained for $a^2R_3(m_l,m_l)$ are collected in Fig.~\ref{fig3c13} and show an almost linear dependence on the light quark mass. 
\vspace{-0.1cm}
Our final determinations
of the mixing coefficient $c_{13}$ are reported in Table~\ref{tb:c13}.
Note that the total uncertainties on $c_{13}$ are of the order of $0.02\% - 0.1\%$. The value which we obtained for $Z_{13}$ in one-loop perturbation theory leads to an estimate for $c_{13}$:
{\small
\be
c_{13}^{pert}= a^2 Z_{13}^{pert} = \frac{g^2\,C_F}{16\,\pi^2}\,\left(36.0613\right)
\ee
}
{\small
\begin{table}
\begin{center}
\begin{tabular}{|l| l|l|}
\hline
$\beta$ & $c_{13}^{pert}$ & $c_{13}$ \\\hline
1.90 & 0.96152 & 0.89769(17)\\\hline
1.95 & 0.93686 & 0.87687(36)\\\hline
2.10 & 0.86995 & 0.81646(78)\\ \hline
\end{tabular}
\caption{Values of the mixing coefficient $c_{13}^{pert}$ obtained at one-loop in perturbation theory and $c_{13}$ obtained as the chiral limit at three values
of the lattice spacing (see Fig.1).} 
\label{tb:c13}
\end{center}
\end{table}
}
It can be seen that our
non-perturbative determinations of $c_{13}$ differ by less than $\sim 10\%$ from the perturbative predictions at one loop. For detailed information on non-perturbative results, see the proceedings by V.~Lubicz in this conference~\cite{VL2014}.

\noindent
{\bf Acknowledgments:} The work of M.~Constantinou and M.~Costa was
supported by the Cyprus Research Promotion Foundation under Contract
No. TECHNOLOGY/${\rm \Theta E\Pi I\Sigma}$/0311(BE)/16.\\
D.~Meloni acknowledges MIUR (Italy) for financial support under the program
Futuro in Ricerca 2010 (RBFR10O36O).\\
We also acknowledge the CPU time provided by the PRACE Research Infrastructure under the project PRA027 at JSC (Germany), and by the agreement between INFN and CINECA under the specific initiative INFN-lqcd123.

\end{document}